\documentstyle[aps, epsfig, floatfig, array, amssymb]{revtex}

\begin{document}

\title{Quantum Logic Using Sympathetically Cooled Ions}

\author{D. Kielpinski, B.E. King, C.J. Myatt, C.A. Sackett,\\ Q.A.
Turchette,
W.M. Itano, C. Monroe, and D.J. Wineland}

\address{Time and Frequency Division\\
National Institute of Standards and Technology\\
Boulder, CO 80303}

\author{W.H. Zurek}

\address{T-6 (Theoretical Astrophysics), MS B288, Los Alamos National
Laboratory, Los Alamos, NM 87545}

\author{Work of the U.S. Government.  Not subject to U.S. copyright.}

\maketitle

\begin{abstract}
One limit to the fidelity of quantum logic operations on trapped ions
arises from
heating of the ions' collective modes of motion.  Sympathetic cooling
of the ions during the logic operations may eliminate this source of
errors.  We
discuss benefits and drawbacks of this proposal, and describe possible
experimental implementations.  We also present an overview of
trapped-ion dynamics in this scheme.
\end{abstract}

\section{Introduction}

One of the most attractive physical systems for generating large
entangled states and realizing a quantum computer \cite{steane} is a
collection of cold trapped atomic ions \cite{cirzol}.  The ion trap
quantum computer stores one or more
quantum bits (qubits) in the internal states of each trapped ion, and
quantum logic gates (implemented by interactions with externally applied
laser beams) can couple qubits through a collective quantized mode of
motion of the ion Coulomb crystal.  Loss of coherence of the
internal states of trapped ions is negligible under proper conditions
but heating of the motion of the ion crystal may ultimately
limit the fidelity of logic gates of this type.  In fact, such
heating is currently a limiting factor in the NIST ion-trap quantum
logic experiments \cite{bible,brian}.\\
\\

Electric fields from the environment readily couple to the motion of the
ions, heating the ion crystal \cite{bible,brian,lam,jheat,henk}.  If the ion
trap is much larger than the ion crystal size, we expect these electric
fields to be nearly uniform across the crystal.  Uniform fields will heat
only modes that involve center-of-mass motion (COM motion),
in which the crystal moves as a rigid body.  Motional modes orthogonal
to the COM motion, for instance the collective breathing mode,
require field gradients to excite their motion.  The heating of these modes
is therefore suppressed \cite{brian}.  However, even
if quantum logic operations use such a ``cold'' mode, the heating of the
COM motion can still indirectly limit the fidelity of logic operations.
Since the laser coupling of an internal qubit and a motional mode depends
on the total wavepacket spread of the ion containing the qubit,
the thermal COM motion can reduce the logic fidelity \cite{bible,brian}.\\
\\

In this paper, we examine sympathetic cooling
\cite{larson} in a particular scheme
for which we can continuously laser-cool the
COM motion while leaving undisturbed the coherences of both the internal
qubits and the mode used for quantum logic.  In this
method, one applies continuous laser cooling to only the center
ion of a Coulomb-coupled string of an odd number of ions.  One can address
the center ion alone if the center ion is of a different ion species
than that composing the rest of the string \cite{fnote1}.
Alternatively, one can simply focus the cooling beams so that they affect
only the center ion.
In either case, the cooling affects only the internal states of the
center ion, leaving all other internal coherences intact.  If the
logic operations use a mode in which the center ion remains at rest,
the motional coherences in that mode are also unaffected by
the cooling.  On the other hand, the sympathetic cooling
keeps the COM motion cold, reducing
the thermal wavepacket spread of the ions.
In the following, we will discuss the dynamics of an ion string in which all
ions are identical except the center ion, assuming heating by a uniform
electric field.  Our results give guidelines for implementing the
sympathetic cooling scheme.  Similar results would apply to two- and
three-dimensional ion crystals \cite{zigzag,walt,drew,travis}.\\
\\

\section{Axial Modes of Motion}

We consider a crystal of $N$ ions, all of charge ${\mathsf q}$, in a
linear RF trap \cite{zigzag,walt}.
The linear RF trap is essentially an RF quadrupole mass
filter with a static confining potential along the filter axis
$\hat{z}$.  If the radial confinement is sufficiently strong compared to
the axial confinement, the ions will line up along the $z$-axis in a
string configuration \cite{zigzag,walt}.  There is no RF electric field along
$\hat{z}$, so we can write the axial
confining potential as $\phi(z)={\mathsf q}a_0z^2/2$ for $a_0$ a constant.
The potential energy of the string is then given by
\begin{equation}
\label{eq:pot}
V(z_1,\ldots z_n) = \frac{1}{2} {\mathsf q} a_0 \sum_{{i=1}}^N z_i^2 +
\frac{{\mathsf q}^2}{8\pi\epsilon_0}
\sum_{\stackrel{i,j=1}{i\not=j}}^N \frac{1}{\mid z_i-z_j\mid}
\end{equation}
for $z_i$ the position of the $i$th ion in the string (counting from the
end of the string).  The first term in the potential energy
expresses the influence of the static confining
potential along the $z$-axis, while the second arises from the mutual
Coulomb repulsion of the ions.  For a single ion of mass $m$, the trap
frequency along $z$ is just $\omega_z=\sqrt{{\mathsf q}a_0/m}$.\\
\\

We can compute the equilibrium positions of the ions in the string by
minimizing the potential energy of Eq. \ref{eq:pot}.
Defining a length scale $\ell$ by $\ell^3 = {\mathsf q}/(4\pi\epsilon_0a_0)$
and normalizing the ion positions by
$u_i = z_i/\ell$ gives a
set of equations for the $u_i$ as
\begin{equation}
u_i - \sum_{j=1}^{i-1}\frac{1}{(u_i-u_j)^2} + \sum_{j=i+1}^{N}
\frac{1}{(u_i-u_j)^2} = 0, \:\: i = 1 \ldots N
\end{equation}
which has analytic solutions only up to $N=3$.  Steane \cite{steane} and
James \cite{james} have
computed the equilibrium positions of ions in strings with $N$ up to 10.
The potential energy is independent of the mass, so the equilibrium
positions of ions in a string are independent of the elemental
composition of the string if all the ions have the same charge.\\
\\

In a real ion trap the ions will have some nonzero temperature and will
move about their equilibrium positions.  If the ions are sufficiently
cold, we can write their positions as a function of time as $z_i(t)=
\ell u_i +
q_i(t)$, where $q_i(t)$ is small enough to allow linearizing all forces.
We specialize to the case of an odd number of ions $N$, where all ions
have mass $m$, except for the one at the center of the string which has
mass $M$.  The ions are numbered $1\ldots N$, with the center ion
labeled by
$n_c=(N+1)/2$.  Following James \cite{james}, the Lagrangian for the
resulting small oscillations is
\begin{eqnarray}
L &=& \frac{m}{2}\sum_{\stackrel{i=1}{i\not=n_c}}^N \dot{q}_i^2 +
\frac{M}{2}\dot{q}_{n_c}^2 - \frac{1}{2}\sum_{i,j=1}^N
\left. \frac{\partial^2 V}{\partial z_i \partial z_j}
\right|_{\{q_i\}=0} q_iq_j\\
&=& \frac{m}{2}\sum_{\stackrel{i=1}{i\not=n_c}}^N \dot{q}_i^2 +
\frac{M}{2}\dot{q}_{n_c}^2 - \frac{1}{2} {\mathsf q}a_0 \sum_{i,j=1}^N
A_{ij}q_iq_j
\end{eqnarray}
where
\begin{equation}
A_{ij} = \left\{ \begin{array}{ll}
{\displaystyle 1+2\sum_{\stackrel{k=1}{k\not=i}}^N\frac{1}{\mid u_i-u_k
\mid^3}} & \mbox{$i=j$}\\
{\displaystyle -2
\frac{1}{\mid u_i-u_j \mid^3}} & \mbox{$i\not=j$} \end{array} \right.
\end{equation}
\\

We define a normalized time as $T=\omega_z t$.
In treating the case of two ion species, we write $\mu=M/m$ for the mass
ratio of the two species and 
normalize the amplitude of the ion vibrations $q_i(t)$ as
$Q_i=q_i\sqrt{{\mathsf q}a_0}$, $i\not=n_c$, $Q_{n_c}=q_{n_c}
\sqrt{{\mathsf q}a_0\mu}$.  The
Lagrangian becomes
\begin{equation}
\label{eq:axl}
L = \frac{1}{2}\sum_{i=1}^N\left(\frac{dQ_i}{dT}\right)^2
-\frac{1}{2}\sum_{i,j=1}^N
A'_{ij}Q_iQ_j
\end{equation}
where
\begin{equation}
A'_{ij} = \left\{ \begin{array}{ll} A_{ij} & \mbox{$i,j \not= n_c$}\\
A_{ij}/\sqrt{\mu} & \mbox{$i \:{\rm or}\: j = n_c,\: i\not=j$}\\
A_{ij}/\mu & \mbox{$i=j=n_c$} \end{array} \right.
\end{equation}
generalizing the result of James \cite{james}.\\

The Lagrangian is now cast in the canonical form for small oscillations
in the coordinates $Q_i(t)$.  To find the normal modes, we solve the
eigenvalue equation
\begin{equation}
{\mathbf A}' \cdot \vec{v}^{(k)} = \zeta_k^2 \vec{v}^{(k)} \hspace{12pt}
k=1\ldots N
\end{equation}
for the frequencies $\zeta_k$ and (orthonormal) eigenvectors $\vec{v}^{(k)}$
of the $N$ normal modes.  Because of our normalization of the Lagrangian
(\ref{eq:axl}), the $\zeta_k$ are normalized to $\omega_z$ and
the $\vec{v}^{(k)}$ are expressed in terms of the normalized coordinates
$Q_i(t)$.  In terms of the physical time $t$, the
frequency of the $k$th mode is $\zeta_k\omega_z$.  If the $k$th mode is
excited with an amplitude $C$, we have
\begin{eqnarray}
q_i(t) &=& {\rm Re}\:
[C v_i^{(k)}e^{i(\zeta_k\omega_zt+\phi_k)}] \hspace{12pt}
i\not=n_c\\
q_{n_c}(t) &=& {\rm Re}\:[C\frac{1}{\sqrt{\mu}} v_{n_c}^{(k)}
e^{i(\zeta_k\omega_zt+\phi_k)}]
\end{eqnarray}
in terms of the physical coordinates $q_i(t)$.\\
\\

We can solve for the normal modes analytically for $N=3$.  Exact expressions
for the normal-mode frequencies are
\begin{eqnarray}
\zeta_1 &=& \left[ \frac{13}{10}+\frac{1}{10\mu}(21-
\sqrt{441-34\mu+169\mu^2}) \right]^{\frac{1}{2}}\\
\zeta_2 &=& \sqrt{3}\\
\zeta_3 &=& \left[ \frac{13}{10}+\frac{1}{10\mu}(21+
\sqrt{441-34\mu+169\mu^2}) \right]^{\frac{1}{2}}
\end{eqnarray}
normalized to $\omega_z$.  The mode eigenvectors are
\begin{eqnarray}
\vec{v}^{(1)} &=& N_1 \left(1, \frac{\sqrt{\mu}}{8}(13-5\zeta_1^2),
1\right)\\
\vec{v}^{(2)} &=& N_2 \:\:(1, 0, -1)\\
\vec{v}^{(3)} &=& N_3 \left(1, \frac{\sqrt{\mu}}{8}(13-5\zeta_3^2), 1\right)
\end{eqnarray}
in terms of $Q_i(t)$.  Here $N_1, N_2,$ and $N_3$ are normalization factors.
In the case of three identical ions ($\mu=1$), we can
express the mode eigenvectors in terms of the $Q_i(t)$ as
$\vec{v}^{(1)}=(1,1,1)/\sqrt{3}$, $\vec{v}^{(2)}=(1,0,-1)/\sqrt{2}$,
and $\vec{v}^{(3)}=(1,-2,1)/\sqrt{6}$.  The mode
eigenvectors, in this special case, also give the ion oscillation amplitudes
in terms of the physical
coordinates $q_i(t)$.  For three identical ions, then, pure axial
COM motion constitutes a normal mode.  (This result holds for an
arbitrary number of identical ions.)  We also note
that the center ion does not move in mode \#2; hence the frequency and
eigenvector of mode \#2 are independent of $\mu$.  For any odd number $N$
of ions there are $(N-1)/2$ modes for which the center ion does not move.
These modes will likewise have frequencies and eigenvectors independent
of $\mu$.  Moreover, they have $v^{(k)}_{n_c-m}=-v^{(k)}_{n_c+m}$ and so
they are orthogonal to the COM motion and do not couple to uniform electric
fields.
The center ion moves in the other $(N+1)/2$ modes, and unless
$\mu=1$, each of these $(N+1)/2$ modes has a component of axial COM
motion and therefore couples to uniform electric fields.\\
\\

For $N = 5$ and higher, the normal mode frequencies depend on $\mu$ in a
complicated way.  However, it is easy to find the
frequencies numerically.  Fig. 1 shows the mode frequencies for $N = 3$,
5, 7, and 9 as a
function of $\mu$ for $0.01<\mu<100$.  The modes are numbered in order of
increasing frequency (at $\mu=1$), and are normalized to $\omega_z$.
In each case, the
lowest-lying mode has all ions moving in the same direction and consists
of pure COM motion for $\mu=1$.  The even-numbered modes correspond to the
$(N-1)/2$ modes for which the center ion does not move.  Their frequencies
are therefore independent of $\mu$.  For both very large and very
small $\mu$ the modes pair up, as shown in Fig. 1.
For each pair there is some value $\mu > 1$ for which the modes become
degenerate.  The relative spacing between modes in a pair
is also smaller in the large-$\mu$ limit than in the small-$\mu$ limit.\\
\\

In selecting a normal mode of motion for logic operations, we want to ensure
that the mode is well resolved from all other normal modes.
However, when modes are nearly degenerate, as for
$\mu \ll 1$ and $\mu \gg 1$, transfer of energy can occur between the
modes in the presence of an appropriate coupling, for instance
if the static confining potential contains small terms of order $z^3$
\cite{bible}.  This coupling can lead to a loss of coherence of the
logic mode.  Also, the need to resolve the logic mode from a nearby
spectator mode can force a reduction in gate speed.
These effects limit the usefulness of the sympathetic cooling scheme for
$\mu$ very large.  Evidently it is best to use a cooling ion
that is of the same mass or lighter than the logic ions.  In this case
mode \#2 is well-separated from all other modes, as shown in Fig. 1.\\
\\

\section{Transverse Modes of Motion}

We now consider the motion of the ions transverse to the $z$-axis.
The ions experience an RF potential
$\chi\cos(\Omega t)(x^2-y^2)/2$
for a suitable choice of axes $x$ and $y$ perpendicular to $z$, where
$\Omega$ is the frequency of the RF field and $\chi$ is a constant.
The static confining potential can be written $({\mathsf q}a_0/2)(z^2-
\alpha x^2-(1-\alpha)y^2)$ at the position of the ions (with $\alpha$ a
constant), so there is also a
transverse static electric field.  To analyze the ion motion, we work in the
pseudopotential approximation, in which one time-averages the motion over a
period of the RF drive to find the ponderomotive force on the ion.
If the static potential is negligible, the RF drive
gives rise to an effective transverse confining potential of
$\frac{1}{2}m\omega_{r0}^2(x^2+y^2)$, where
$\omega_{r0}={\mathsf q}\chi/(\sqrt{2}\Omega m)$
for an ion of mass $m$.  If we include the effects of the static field,
the transverse potential becomes
$\frac{1}{2}m(\omega_x^2x^2+\omega_y^2y^2)$, where
$\omega_x=\omega_{r0}\sqrt{1-\alpha\omega_z^2/\omega_{r0}^2}$,
$\omega_y=\omega_{r0}\sqrt{1-(1-\alpha)\omega_z^2/\omega_{r0}^2}$.  Below
we will assume $\alpha=1/2$, so that $\omega_y=\omega_x$.
In any case,
the transverse potential is that of a simple harmonic oscillator, as we
saw also for the axial potential.  However, the transverse potential
depends directly on the ion's mass, so the center ion of a string
feels a different trap potential than the others for $\mu\not=1$.\\
\\

We define $\epsilon=\omega_{r0}/\omega_z$, so that
$\omega_x=\omega_z\sqrt{\epsilon^2-1/2}$.
Then the normalized Lagrangian for the motion along $x$ is
\begin{equation}
\label{eq:trans}
L = \frac{1}{2}\sum_{i=1}^N\left(\frac{dX_i}{dT}\right)^2
- \frac{1}{2}\sum_{i,j=1}^N
B'_{ij}X_iX_j
\end{equation}
where $X_i = x_i\sqrt{{\mathsf q}a_0}$ for $i \not= n_c$ and $X_{n_c} =
x_i\sqrt{{\mathsf q}a_0\mu}$ are normalized ion vibration amplitudes
along $x$.  Here
\begin{equation}
B'_{ij} = \left\{ \begin{array}{ll} B_{ij} & i,j \not= n_c\\
B_{ij}/\sqrt{\mu} & i \:{\rm or}\: j = n_c,\: i\not=j\\
B_{ij}/\mu & i=j=n_c \end{array} \right.\\
\end{equation}
and
\begin{equation}
B_{ij} = \left\{ \begin{array}{ll} {\displaystyle
\epsilon^2-\frac{1}{2}
-\sum_{\stackrel{k=1}{k\not=i}}^N\frac{1}{\mid u_i-u_k \mid^3}}
& \mbox{$i=j, j\not=n_c$}\\
{\displaystyle
\frac{\epsilon^2}{\mu}-\frac{1}{2}
-\sum_{\stackrel{k=1}{k\not=i}}^N\frac{1}{\mid u_i-u_k \mid^3}}
& \mbox{$i=j=n_c$}\\
{\displaystyle
\frac{1}{\mid u_i-u_j \mid^3}} & \mbox{$i\not=j$}\\
\end{array} \right.
\end{equation}
We can describe the normal mode frequencies and oscillation amplitudes
in terms of the eigenvectors and eigenvalues of $B'_{ij}$, just as
for the axial case above.  The normalizations of the time and position
coordinates remain the same as in the axial case.\\
\\

In the previous section, we assumed that the radial confinement of the
ions was strong enough that the configuration of ions in a string along
the $z$-axis was always stable.  However, for
sufficiently small $\epsilon$, the string configuration becomes
unstable.  The stable configurations for different values of $\epsilon$ can
be calculated \cite{rob,schiff}, and several of these configurations
have been
observed for small numbers of ions \cite{zigzag,walt}.  Rather than review
the theory of these configurations, we will simply find the range of
validity of our small-oscillation Lagrangian for the string configuration.
The string will remain stable for all $\epsilon$ greater than some
$\epsilon_s=\epsilon_s(\mu)$; $\epsilon_s$ also varies with $N$.
On the boundary between stable and unstable
regions, the frequency of some mode goes to zero.  Recalling that the
determinant of a matrix is equal to the product of its eigenvalues, we
see that $\epsilon_s(\mu)$ is the maximum value of $\epsilon$ satisfying
 $\det B'(\epsilon, \mu)=0$ for $\mu$ fixed.  Fig. 2 shows $\epsilon_s(\mu)$
as a function of $\mu$ for 3, 5, 7, and 9 ions.  In each case, there is
a cusp in $\epsilon_s(\mu)$ corresponding to the crossing of the two largest
solutions to $\det B'(\epsilon, \mu)=0$.  The position of the cusp varies
with the number of ions, but lies between $\mu=0.1$ and
$\mu=1$ for $N\leq 9$.  Only the cusp for $N=3$ is clearly visible in
Fig. 2, but numerical study indicates the presence of a cusp for all
four values of $N$.  For $\mu$ greater than the value at the cusp,
$\epsilon<\epsilon_s(\mu)$ corresponds to instability of the
zigzag mode, so that the string breaks into
a configuration in which each ion is displaced in the opposite direction to
its neighbors \cite{rob,schiff}.  For
$\mu$ smaller than the value at the cusp, $\epsilon_s$ is independent of
$\mu$.  In this regime, $\epsilon<\epsilon_s$ creates an
instability in a mode similar to the zigzag mode, except that the center
ion remains fixed.\\
\\

We can proceed to calculate the frequencies of the transverse
modes for values $\epsilon>\epsilon_s(\mu)$.  Again, these frequencies are
normalized to the axial frequency of a single ion of mass $m$.
Fig. 3 shows the transverse mode
frequencies for 3, 5, 7, and 9 ions as a function of $\mu$, where $\epsilon$
is taken equal to $1.1\epsilon_s(\mu)$.  The modes are numbered in order of
increasing frequency
at $\mu=1$ (all ions identical).  In this numbering scheme, the central
ion moves in odd-numbered modes but not in even-numbered modes.  The
frequencies of the even-numbered modes appear to depend on $\mu$ because
they are calculated at a multiple of $\epsilon_s(\mu)$; for constant
$\epsilon$ these frequencies are independent of $\mu$.  The cusps
in the mode frequencies in
Fig. 3 arise from the cusps of $\epsilon_s(\mu)$ at the crossover points
between the two relevant
solutions of $\det B'=0$.  Mode frequencies plotted
for a constant value of $\epsilon$ do not exhibit these cusps.  As in the case
of axial motion, the mode frequencies form pairs of one even- and one
odd-numbered mode for small $\mu$.  However, for large
$\mu$ all but one of the transverse modes become degenerate.  The only
nondegenerate transverse mode in this case is the zigzag mode.  In general,
the modes are most easily resolved from their neighbors for $\mu=1$,
as in the case of axial motion.  Increasing $\epsilon$ reduces the
frequency spacing between nearly degenerate modes.  At
$\epsilon=1.1\epsilon_s(\mu)$ and $\mu=1$, for instance,
the fractional spacing between the cold transverse mode of 3 ions and its
nearest neighbor is 0.20, but for
$\epsilon=1.5\epsilon_s(\mu)$ the same spacing is 0.09.\\
\\

The near-degeneracy of the modes
for large or small $\mu$ and for $\epsilon/\epsilon_s$ significantly greater
than 1 limits the usefulness of these modes because of possible
mode cross-coupling,
just as for the axial modes.  Resolving a particular transverse
mode requires operating the trap near the point at which the string
configuration becomes unstable, i.e., $\epsilon$ near $\epsilon_0(\mu)$.
In this regime, the collective motion
of the ions is quite sensitive to uncontrolled perturbations, which may
pose significant technical problems for using a transverse mode in quantum
logic operations.\\
\\

\section{Mode Heating}

Stochastic electric fields present on the ion trap electrodes, for instance
from fluctuating surface potentials, can heat the various normal modes of
motion incoherently. For ion trap characteristic dimension
$d_{trap}$ much larger than
the size of the ion crystal $d_{ions}$, these fields are approximately
uniform across the ion crystal, so they couple only to the COM motion.
The $(N-1)/2$ even-numbered modes are orthogonal to the COM motion,
so they are only heated by
fluctuating electric field gradients.  The heating rates of these
modes are reduced by a factor of at least
$(d_{ions}/d_{trap})^2 \ll 1$ as compared to the heating of the other
modes \cite{brian}.  In the following, therefore, we will neglect the effects
of fluctuating field gradients, so that the even-numbered modes do not heat
at all.\\
\\

The analysis of sections 2 and 3 shows that the motion of a crystal of
$N$ ions is separable into the $3N$ normal modes, each of which is
equivalent to a simple harmonic oscillator.  Hence we can quantize the
crystal motion by quantizing the normal modes.  The $k$th normal mode gives
rise to a ladder of energy levels spaced by $\hbar\zeta_k\omega_z$, with $3N$
such ladders in all.
If we now write the uniform electric field power
spectral density as $S_E(\omega)$, we can generalize the result of
\cite{forts} to give
\begin{equation}
\label{eq:heat}
\dot{\overline{n}}_k =
\frac{{\mathsf q}^2 S_{E}(\zeta_k\omega_z)}{4m\hbar\zeta_k\omega_z}
\left(
\frac{v^{(k)}_{n_c}}{\sqrt{\mu}}+\sum_{\stackrel{j=1}{j\not=n_c}}^N
v^{(k)}_{j} \right)^2
\end{equation}
for the heating rate of the $k$th mode, expressed in terms of the average
number of quanta gained per second.  Recall that $v^{(k)}_i$ is the
oscillation amplitude of the $i$th ion in the $k$th normal mode,
expressed in the normalized coordinates.
It is useful to normalize the heating rate
in equation (\ref{eq:heat}) to the heating rate of the lowest-lying
axial mode of a string of identical ions.  This normal mode consists
entirely of COM motion and we write $v^{COM}_j=1/\sqrt{N}$ for all ions.
The normalized heating rate of the $k$th mode is then
\begin{equation}
\label{eq:hnorm}
\frac{\dot{\overline{n}}_k}{\dot{\overline{n}}_{COM}} =
\frac{1}{N\zeta_k} \left(
\frac{v^{(k)}_{n_c}}{\sqrt{\mu}}+\sum_{\stackrel{j=1}{j\not=n_c}}^N
v^{(k)}_{j} \right)^2
\end{equation}
where we have assumed that the spectral density $S_E(\omega)$ is
constant over the frequency range of the normal modes, i.e.,
$S_E(\omega_z)=S_E(\zeta_k\omega_z)$.\\
\\

Fig. 4 shows plots of the normalized heating rates of the axial modes
for $N = 3$, 5, 7, and 9 as a function of $\mu$.
Fig. 5 is the same, but for the transverse modes, with
$\epsilon=1.1\epsilon_s$.  The numbering of modes on
the plots of heating rate matches the numbering on the corresponding
plots of mode frequency (Figs. 1 and 2).\\
\\

In both axial-mode and transverse-mode plots, the even-numbered modes have
the center ion at rest, while the center ion moves for all
odd-numbered modes.  We see from Figs. 4 and 5 that the modes for which
the center ion is fixed can never heat, while all the other modes
always heat to some extent for $\mu\not=1$.  We will refer to these modes
as ``cold" and ``hot" modes, respectively.  If the ions are identical, only
the modes with all ions moving with the same amplitude (COM modes) can heat.
There are three such modes, one
along $\hat{x}$, one along $\hat{y}$, and one along $\hat{z}$.  In
interpreting Figs. 4 and 5, it is important to recall that the
normalized heating rate defined in Eq. (\ref{eq:hnorm}) is inversely
proportional to the mode frequency.  For instance, the $\mu$-dependence of
the heating rate of the highest-frequency transverse mode can be largely
ascribed to variations in the mode frequency, rather than to
changes in the coupling of the mode to the electric field.\\
\\

\section{Prospects for Sympathetic Cooling}

Heating reduces logic gate fidelity in two ways.  The logic mode
itself can be heated, but by choosing a cold mode, we can render this effect
negligible.  On the other hand, the Rabi frequency of the transition
between logic-mode motional states depends on the total wavepacket
spread of the ion involved in the transition \cite{bible,brian}.  Heating on
modes other than the logic mode can thus lead to unknown, uncontrolled
changes in this Rabi frequency, resulting in overdriving or underdriving
of the transition.  The purpose of sympathetic cooling is to remove this
effect by cooling the center ion and thus all hot modes.\\
\\

For sympathetic cooling to be useful,
we must find a cold mode suitable for use in quantum logic.  The
cold mode must be spectrally well separated from any other modes in order
to prevent
unwanted mode cross-coupling.  We can use the lowest-lying cold axial mode
as the
logic mode for $\mu \lesssim 3$.  In this mode, called the breathing mode,
the center ion remains fixed and the spacings between ions expand
and contract in unison.  Unless the trap is operated very close
to the instability point of the string configuration, the breathing mode is
better separated from its neighbors than are any of the cold transverse
modes.  For $\mu \gtrsim 3$ any cold mode, either
axial or transverse, is nearly degenerate with a hot mode.  In this regime
one must make a specific calculation of mode frequencies in order to
find the best-resolved cold mode.  Even so, the cold axial modes
are again better separated from their neighbors than are the cold
transverse modes, except for $\epsilon$ very close to $\epsilon_s(\mu)$.
It seems best to select a cold axial mode as the logic mode in most cases.\\
\\

By selecting our laser-beam geometry appropriately, we can ensure that
the Rabi frequency of the motional transition on the axial
mode used for logic depends chiefly on the spread
of the ion wavepacket along $z$.  In this case, heating of the
axial modes will affect logic-gate fidelity, but heating of the transverse
modes will have little effect.  If the mass of the central ion is nearly
the same as that of the others ($\mu\approx1$), only the lowest
axial mode will heat significantly, and we can continuously cool
this mode by cooling only the central ion, ensuring that all ions remain in
the Lamb-Dicke limit \cite{bible}.
If $\mu$ is not near 1, we must cool all $(N+1)/2$ hot modes (again by
addressing the central ion) to keep all ions in the Lamb-Dicke limit.\\
\\

The analysis above indicates that, all other things being equal, we are
best off if our substituted ion is identical to, or is an isotope of,
the logic ions.  However,
sympathetic cooling can still be useful if the two ion species have
different masses.  For example, we can consider sympathetic cooling
using the species ${}^9$Be${}^+$ and ${}^{24}$Mg$^+$.  Linear traps
constructed at
NIST have demonstrated axial secular frequencies of over 10 MHz for
single trapped ${}^9$Be${}^+$ ions.  For three ions with
${}^{24}$Mg$^+$
as the central ion, $\omega_z({\rm Be}^+)=2\pi\times10$ MHz yields a spacing
of 1.6 MHz between the cold axial breathing mode and its nearest
neighbor.  If we reverse the roles of the ions
($\omega_z({\rm Mg}^+)=2\pi\times10$ MHz), the spacing increases to
6.2 MHz.  The transverse modes are much harder to resolve from
each other.  For three ions with ${}^{24}$Mg$^+$ in the center, we
require $\omega_{r0}({\rm Be}^+)=2\pi\times 27.6$ MHz to obtain
$\epsilon=1.1\epsilon_s$, and the
spacing between the cold transverse zigzag mode and its nearest neighbor
is only 560 kHz.  Reversing the roles of the ions, we find
$\epsilon=1.1\epsilon_s$ at $\omega_{r0}({\rm Mg}^+)=2\pi\times 14.7$ MHz
with a spacing
of 1.1 MHz.  For this combination of ion species, the cold axial breathing
mode seems most appropriate for logic.
For a string of 3 or 5 ions,
sympathetic cooling would require driving transitions on 2 or 3
axial-mode sidebands, respectively.  From this example we see that
sympathetic cooling can be useful even for ion mass ratios of nearly
3 to 1.\\
\\

\section{Conclusion}

We have investigated a particular sympathetic cooling scheme for
the case of an ion string confined in a linear RF trap.  We have
numerically calculated the mode frequencies of the axial
and transverse modes as functions of the mass ratio $\mu$ and trap
anisotropy $\epsilon$ for 3, 5, 7, and 9 ions.  We have also calculated the
heating rates of these modes relative to the heating rate of a single ion,
assuming that the heating is driven by a uniform stochastic electric field.
The results indicate that the scheme is feasible for many choices of ion
species if we use a cold axial mode as the logic mode.  The optimal
implementation of the scheme employs two ion species of nearly
equal mass.
However, a demonstration of sympathetic cooling using ${}^9$Be${}^+$ and
${}^{24}$Mg$^+$ appears well within the reach of
current experimental technique.\\
\\

\acknowledgements{This research was supported by NSA, ONR, and ARO.  This
publication is the work of the U.S. Government and is not subject to
U.S. copyright.}

\newpage

\begin{figure}[top]
\vspace{2 cm}
\epsfig{file=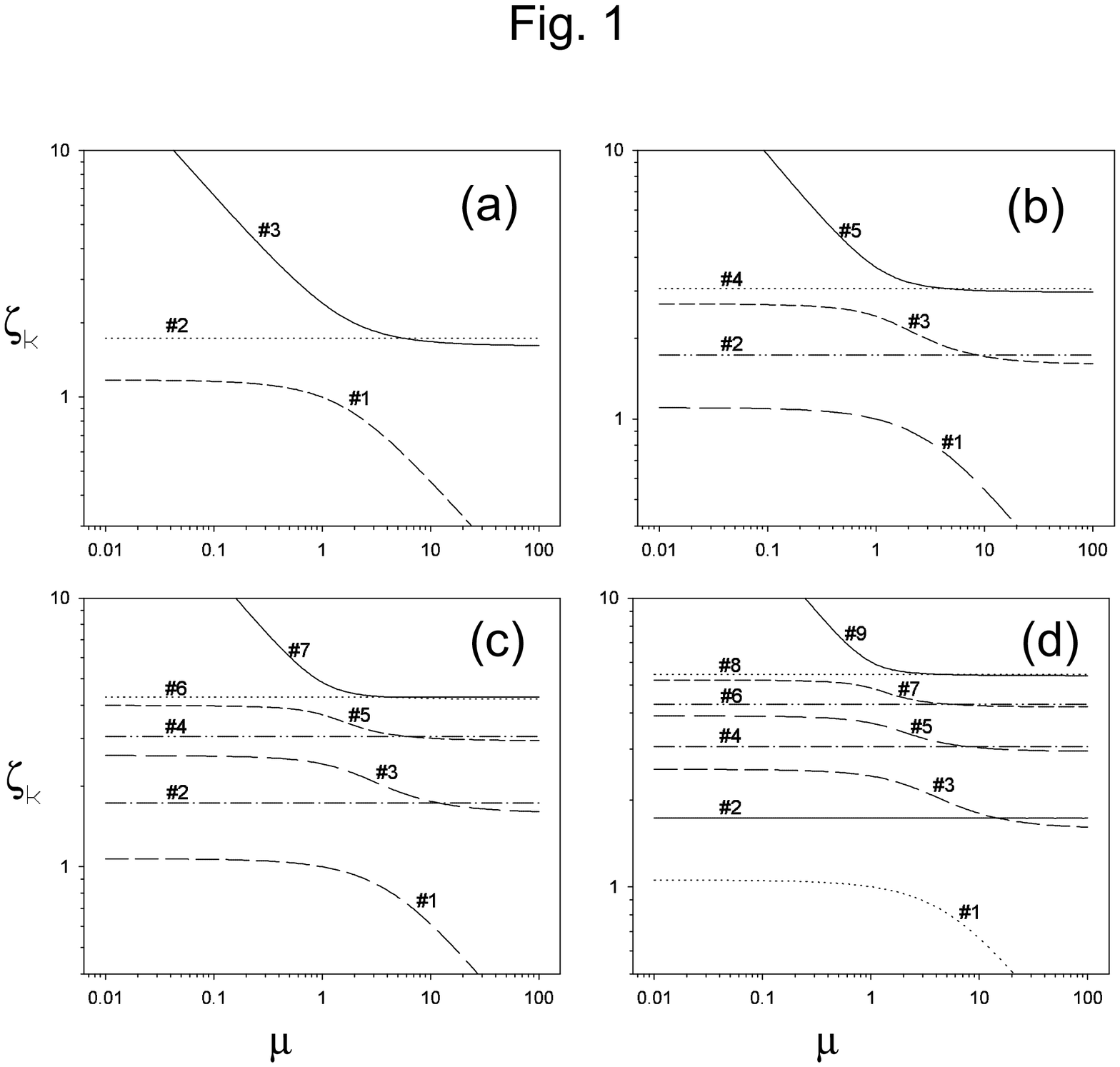, width=440pt}
\caption{Normalized axial mode frequencies as a function of $\mu$
for (a) 3, (b) 5, (c) 7, and (d) 9 ions.}
\end{figure}

\newpage

\begin{figure}[top]
\epsfig{file=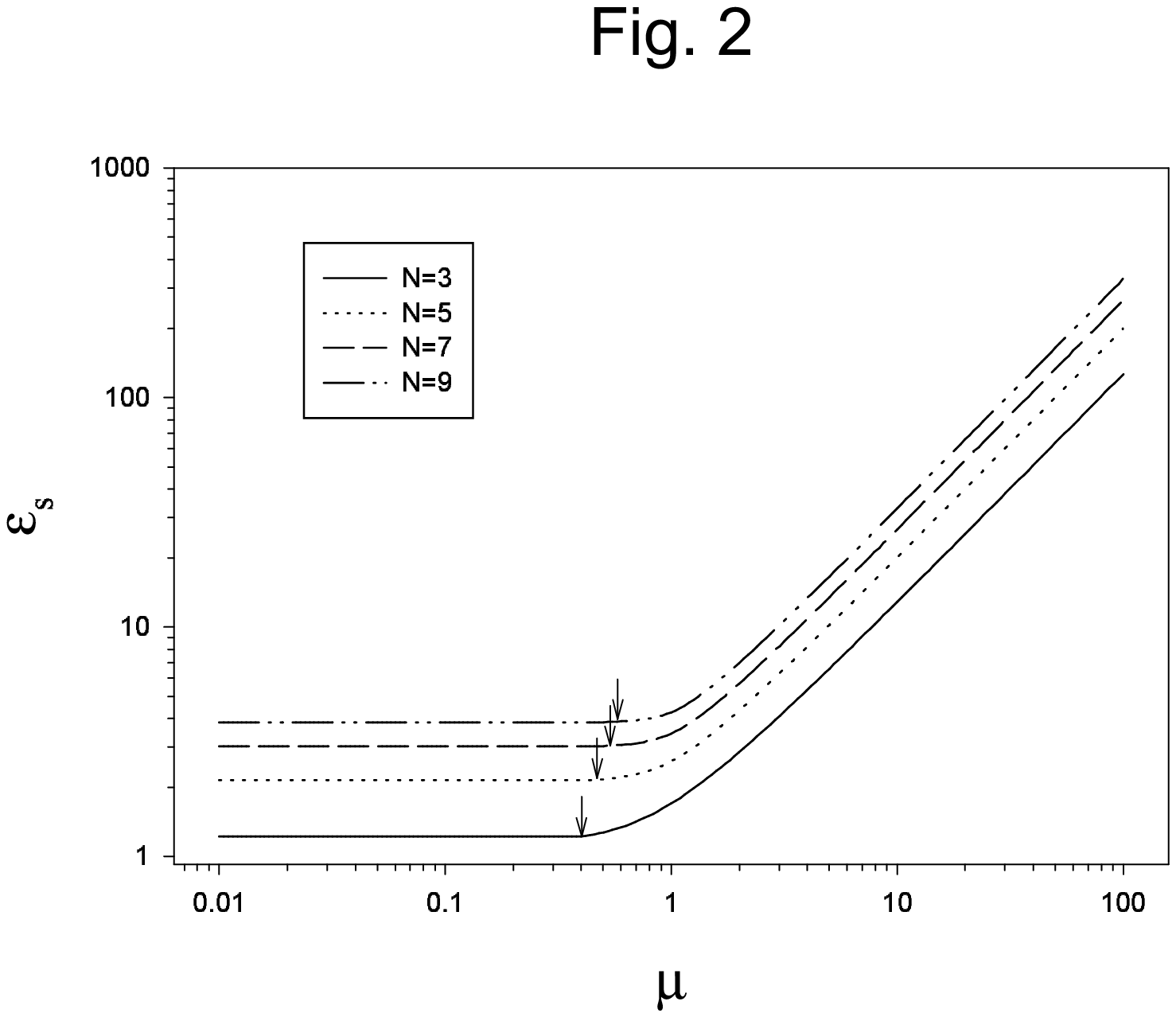, width=440pt}
\caption{Trap anisotropy at instability of the string configuration
as a function of
$\mu$ for 3, 5, 7, and 9 ions.  Arrows indicate the cusps discussed in
the text.}
\end{figure}

\newpage

\begin{figure}[top]
\epsfig{file=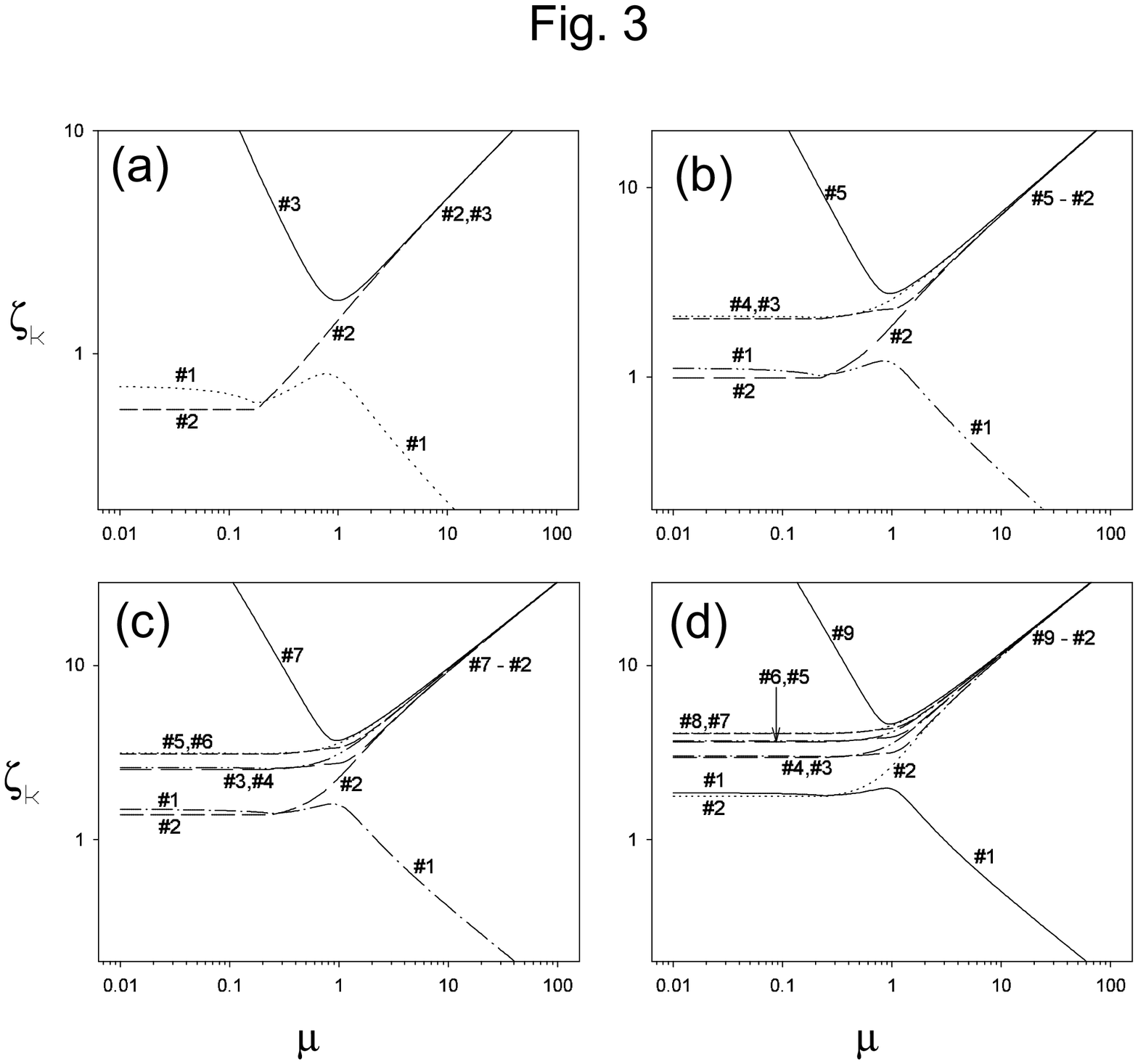, width=440pt}
\caption{Normalized frequencies of the transverse modes as a function of
$\mu$ with $\epsilon=1.1\epsilon_0(\mu)$ for (a) 3, (b) 5, (c) 7, and (d) 9
ions.}
\end{figure}

\newpage
\begin{figure}[top]
\epsfig{file=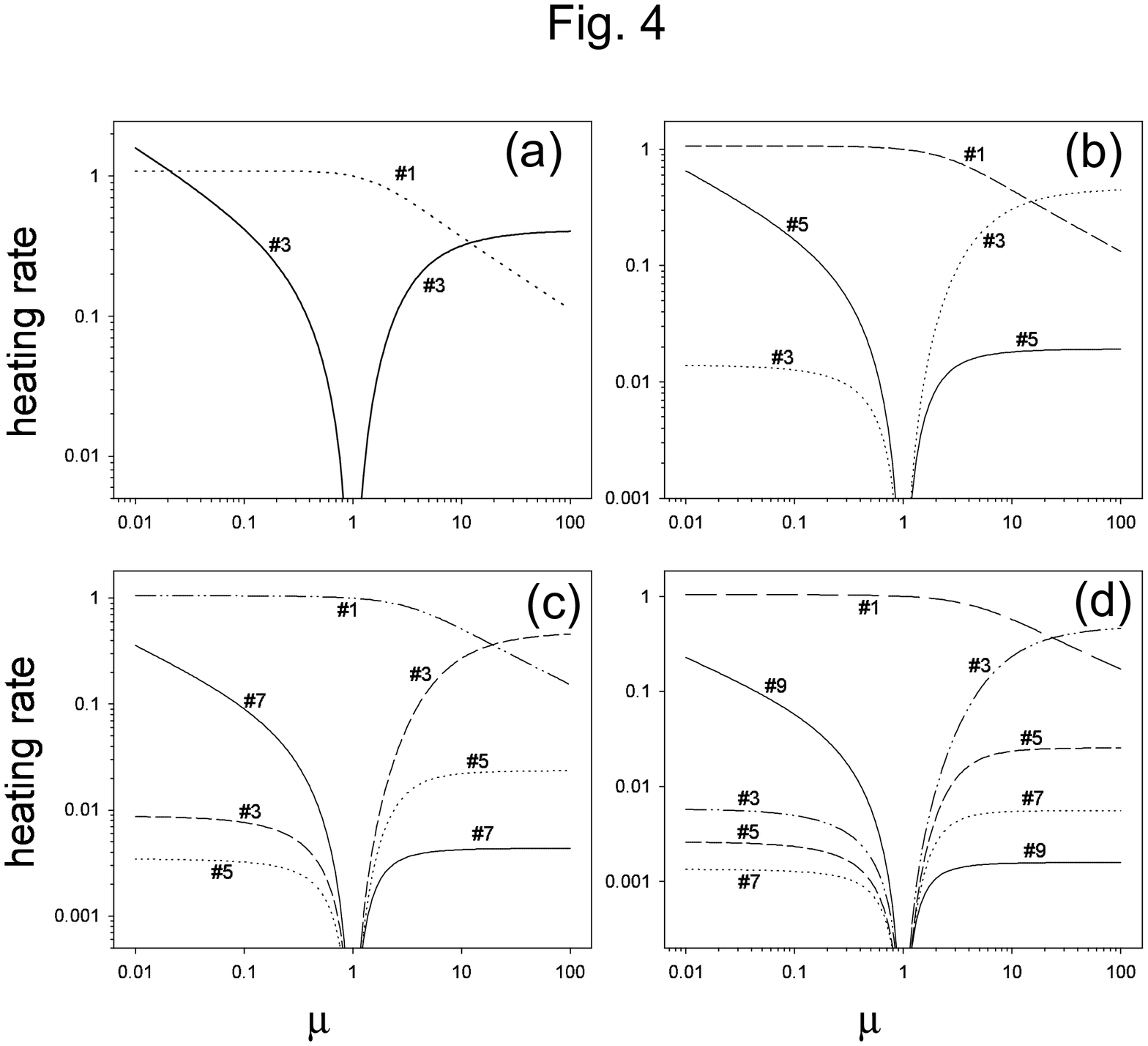, width=440pt}
\caption{Normalized heating rates for the axial modes as a
function of $\mu$ for (a) 3, (b) 5, (c) 7, and (d) 9 ions.}
\end{figure}

\newpage
\begin{figure}[top]
\epsfig{file=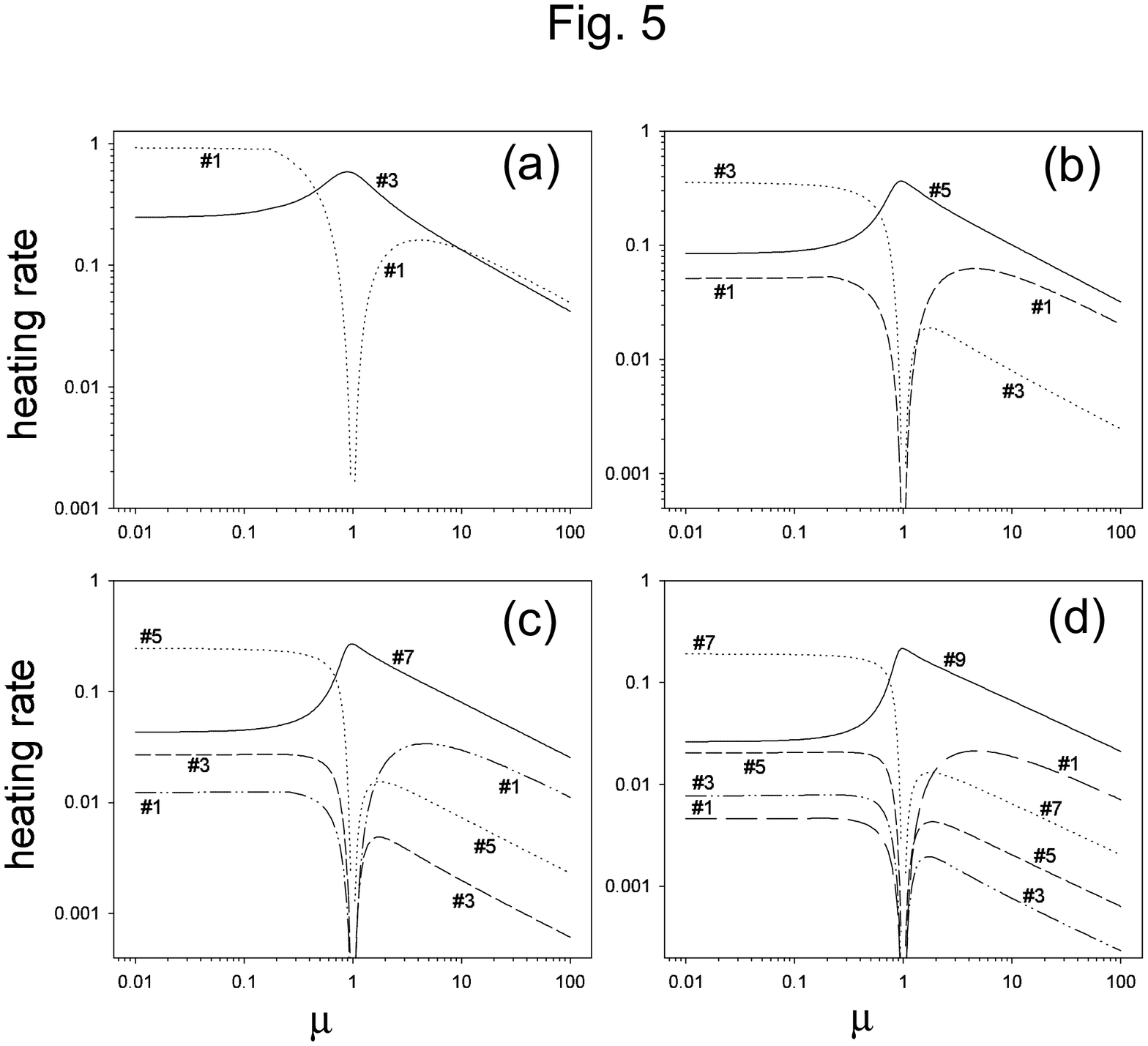, width=440pt}
\caption{Normalized heating rates for the transverse modes as a
function of $\mu$ with $\epsilon=1.1\epsilon_0(\mu)$
for (a) 3, (b) 5, (c) 7, and (d) 9 ions.}
\end{figure}


\begin{references}
\bibitem{fnote1}  We assume that the cooling transitions of the center ion
are well-resolved from all relevant transitions of the logic ions.
\end{references}
\end{document}